\documentclass[aps,pra,showpacs,twocolumn,superscriptaddress,a4paper,amsmath,amssymb]{revtex4}

\usepackage{graphicx}

\usepackage{ulem}
\usepackage[usenames]{color}

\begin{document}

\title{Dissipation-induced squeezing
}

\author{Gentaro Watanabe}
\affiliation{Asia Pacific Center for Theoretical Physics (APCTP),
Pohang, Gyeongbuk 790-784, Korea}
\affiliation{Department of Physics, POSTECH,
Pohang, Gyeongbuk 790-784, Korea}
\affiliation{Nishina Center, RIKEN, 2-1 Hirosawa, Wako, 
Saitama 351-0198, Japan}
\author{Harri M\"akel\"a}
\affiliation{Department of Physics, Ume\aa \,\,University, SE-90187 Ume\aa, Sweden}


\date{\today}

\begin{abstract}
We present a method for phase and number squeezing in
two-mode Bose systems using dissipation.  The effectiveness of this
method is demonstrated by considering cold Bose gases trapped in a
double-well potential. The extension of our formalism to an optical
lattice gives control of the phase boundaries of the steady-state
phase diagram, and we discover a new phase characterized by a non-zero
condensate fraction and thermal-like particle-number statistics.  We
also show how to perform amplitude squeezing 
in a single-mode system using dissipation.
\end{abstract}


\maketitle

\section{Introduction}

In the usual situations, dissipation, caused by coupling to the
environment, is considered to be a serious enemy to quantum-mechanical
systems as it leads to a rapid decay of the
coherence.  Surprisingly, however, an appropriately designed coupling
between the system and the reservoir can drive the system into a given
pure state \cite{Diehl08,Kraus08,Verstraete09}.  
This type of quantum-state engineering, driven by
dissipation, recently has attracted considerable interest both theoretically
\cite{Diehl08,Kraus08,Witthaut08,GarciaRipoll09,Verstraete09,Diehl10,Weimer10,Diehl10b,Tomadin10} 
and experimentally \cite{Krauter10,Barreiro10}.  
A strong advantage of this approach is that 
the desired steady state is obtained without 
active control of the system. This should be contrasted with the standard 
approach in quantum-state engineering where dynamical control 
of the system is required (see, e.g., Ref.\ \cite{dynamical}).
Another advantage is that the target state can be obtained regardless of the initial state,   
making the state-engineering protocol insensitive to imperfections in the initial-state preparation.

In this paper, we present a method for phase and number squeezing
in two-state Bose systems using dissipation. Our work is motivated by the  
 importance of squeezed states in matter-wave interferometry 
\cite{Esteve08,Appel09,Cronin09,Leroux10,Gross10,Grond10,Lee11}. By using squeezed states, 
the performance of an interferometer can be increased:  
Phase-squeezed states improve the accuracy of the readout of the phase difference, and number-squeezed states make longer measurement times possible. 
The scheme presented here gives a way to 
perform both  phase and number squeezing in cold atomic gases and
provides an important building block
for dissipation-driven quantum-state engineering. Additionally, 
by extending our scheme to an optical lattice, we 
demonstrate that it can produce a new non-equilibrium phase
characterized by a non-zero condensate fraction and 
thermal-like particle number statistics.

Our method can be applied to any two-state Bose system, such as   
the one described by the two-site Bose-Hubbard Hamiltonian
$\hat{H}_{\rm BH} =  -2J\hat{S}_x + U \hat{S}_z^2$. 
This Hamiltonian is often used
to describe cold Bose gases trapped in a double-well potential.
Here, $J$ is the tunneling matrix element, $U$ is the on-site interaction, and
we have introduced the SU(2) generators defined as
$\hat{S}_x=(\hat{a}_1^\dag\hat{a}_2+\hat{a}_2^\dag\hat{a}_1)/2$,
$\hat{S}_y=-i(\hat{a}_1^\dag\hat{a}_2-\hat{a}_2^\dag\hat{a}_1)/2$, and
$\hat{S}_z=(\hat{a}_1^\dag\hat{a}_1-\hat{a}_2^\dag\hat{a}_2)/2$,
where $\hat{a}_i (\hat{a}_i^\dag)$ annihilates (creates) an atom in mode $i$.

\section{Two-mode Squeezing\label{2mode}}

First, we consider a two-state Bose system 
with a fixed number of particles.
We assume that the system is coupled to an environment, leading
to dissipative dynamics such that the time evolution of the density operator
is governed by the master equation,
\begin{equation}
  \frac{\partial \hat{\rho}}{\partial t}
= -i[\hat{H},\hat{\rho}] 
+ \frac{\gamma}{2} \left(2\hat{c}\hat{\rho}\hat{c}^\dagger
- \hat{c}^\dagger\hat{c}\hat{\rho} - \hat{\rho}\hat{c}^\dagger \hat{c}
\right),
\label{eq_master}
\end{equation}
where $\hat{c}$ is the Lindblad, or jump, operator. 
We propose the following jump operator 
for the creation of  phase- and number-squeezed states 
(hereafter called the squeezing jump operator):
\begin{align}
  \hat{c}\equiv& (\hat{a}_1^\dagger + \hat{a}_2^\dagger) (\hat{a}_1-\hat{a}_2)
+ \epsilon (\hat{a}_1^\dagger - \hat{a}_2^\dagger) (\hat{a}_1+\hat{a}_2)
\nonumber\\
=& 2 (1+\epsilon) \hat{S}_z - 2i (1-\epsilon) \hat{S}_y.
\label{eq_jump}
\end{align}
Here, $\epsilon$ $(-1<\epsilon <1)$ is a parameter by which we can control the 
squeezing \cite{note_nonzerophi}. 
The jump operator \eqref{eq_jump} can be realized 
in a system of trapped ultracold Bose gas immersed in a background 
Bose-Einstein condensate of a different species of bosonic atoms 
\cite{squeeze2,note_implement}.

In order to understand the action of our squeezing jump operator, we first
consider  the ideal case where there is no Hamiltonian and the
dynamics is driven by the dissipative terms alone. In the double-well 
setting, for example, this can be achieved 
by making the potential barrier between the wells high enough so that $J=0$. 
The interaction $U$ can be reduced using Feshbach resonances.
We start by calculating the amount of squeezing in the steady state.
It can be characterized 
by the normalized phase- and number-squeezing parameters 
\begin{equation}
\xi_{\rm P}^2 \equiv \frac{2\langle\Delta\hat{S}_y^2\rangle}{|\langle\hat{S}_x\rangle|},\quad 
\xi_{\rm N}^2\equiv \frac{2\langle \Delta \hat{S}_z^2\rangle}{|\langle\hat{S}_x\rangle|}, 
\end{equation}
respectively, where $\langle\Delta\hat{S}_{y,z}^2\rangle 
\equiv \langle\hat{S}_{y,z}^2\rangle - \langle\hat{S}_{y,z}\rangle^2$
\cite{Kitagawa93}. 
In this work, the average spin is always parallel to the $x$ axis, 
$\langle\hat{\mathbf{S}}\rangle=\langle\hat{S}_x\rangle\, \hat{\mathbf{e}}_x$. 
Using the coherent-state approximation 
$\langle \hat{S}_x^2\rangle \simeq \langle \hat{S}_x\rangle^2$
with $\langle \hat{S}_x\rangle \simeq N/2 + O(N^0)$, 
which holds for any value of $\epsilon$ provided 
the number of particles $N\gg 1$ \cite{note_approx}
and the truncation scheme based on the Bogoliubov backreaction formalism
$
\langle \hat{S}_i \hat{S}_j \hat{S}_k\rangle
\simeq \langle \hat{S}_i\hat{S}_j\rangle \langle \hat{S}_k\rangle
+ \langle \hat{S}_i\rangle \langle \hat{S}_j\hat{S}_k\rangle 
+ \langle \hat{S}_i\hat{S}_k\rangle \langle \hat{S}_j\rangle
-2 \langle \hat{S}_i\rangle \langle \hat{S}_j\rangle \langle \hat{S}_k\rangle
$ \cite{Vardi01,note_approx,note_bbr}, 
the equations of motion for $\langle \hat{S}_y^2\rangle$ and 
$\langle \hat{S}_z^2\rangle$ become
\begin{equation}
\frac{d}{dt}\langle \hat{S}_{y,z}^2\rangle \simeq
-4N\gamma (1-\epsilon^2) \langle\hat{S}_{y,z}^2\rangle 
+ N^2 \gamma (1\pm\epsilon)^2.
\label{eq_eom}
\end{equation}
In the second term on the right-hand side, the upper sign corresponds 
to $y$, and the lower one corresponds to $z$.
The time constant $\tau$ of $\langle \hat{S}_y^2\rangle$ and 
$\langle \hat{S}_z^2\rangle$, thus, is 
$\tau^{-1} \simeq 4N\gamma (1-\epsilon^2)$.
Since $\langle \hat{S}_y\rangle=\langle \hat{S}_z\rangle=0$ throughout
the time evolution, in the steady state, we get 
\begin{equation}
\langle\Delta\hat{S}_{y,z}^2\rangle = \langle \hat{S}_{y,z}^2\rangle 
\simeq \frac{N}{4} \frac{1\pm\epsilon}{1\mp\epsilon},
\label{Syvariance}
\end{equation}
where, again, the upper sign corresponds to $y$ and the lower sign corresponds to $z$. 
We see that
a phase-squeezed state characterized by $\xi_{\rm P}<1$ is obtained 
for $\epsilon<0$, while $\epsilon>0$ yields a number-squeezed state 
for which $\xi_{\rm N}<1$.

\begin{figure}[t!]
\resizebox{8.2cm}{!}
{\includegraphics{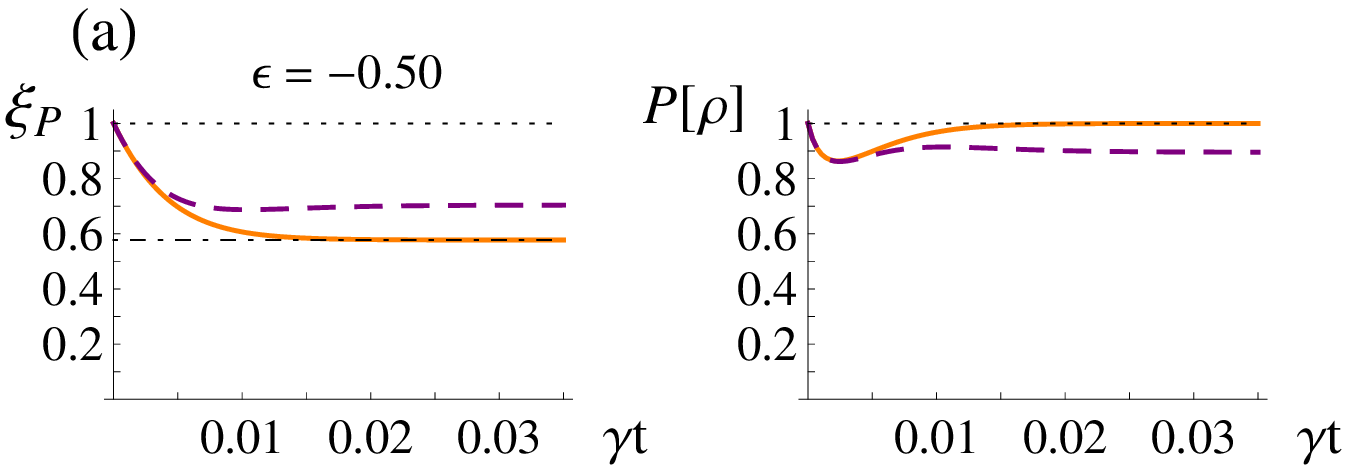}}\\
\resizebox{8.2cm}{!}
{\includegraphics{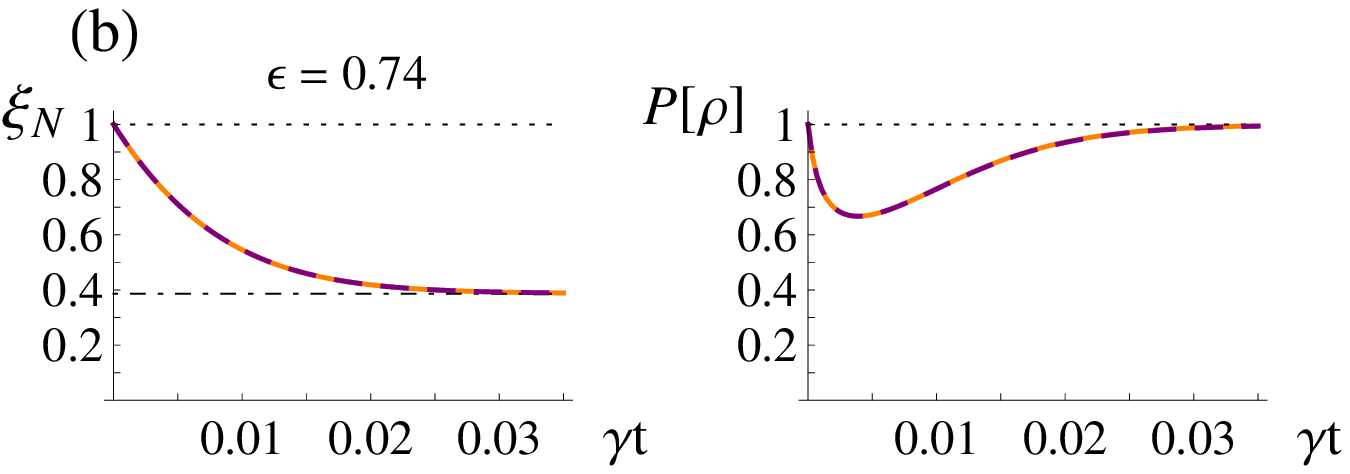}}
\caption{\label{figure1}(Color online)
Time evolution of the squeezing and purity 
for $N=100$, $J=0$ and (a) $\epsilon =-0.50$, (b) $\epsilon =0.74$.  
The solid (dashed) line shows the numerically obtained exact solution for 
$U/\gamma=0$ ($U/\gamma=0.5$), and 
the dashed-dotted line gives the analytical value 
of $\xi_{\rm P,N}$ evaluated using Eq.\ \eqref{Syvariance}.
The initial state is the coherent state $\phi_{\rm sq}|_{\epsilon=0}$.
}
\end{figure}

In general, the steady state of the time evolution is a mixed state.
The amount of mixedness can be quantified using purity 
$P[\hat{\rho}]\equiv\{(N+1)\textrm{Tr}[\hat{\rho}^2]-1\}/N$. 
For a pure state,
$P[\hat{\rho}]=1$, while the completely mixed state gives
$P[\hat{\rho}]=0$. 
The solid lines in Fig. \ref{figure1} show the time evolution 
of the squeezing and purity for $N=100$ (and $J=U=0$) obtained 
by numerically solving the master equation \eqref{eq_master} with 
the squeezing jump operator \eqref{eq_jump}. 
We have set $\epsilon=-0.5$ (upper panels) and 0.74 (lower panels).
The value $\epsilon=0.74$ is chosen to demonstrate that our method 
can produce the same (and even larger) amount of squeezing 
than a different method used in an experiment with cold atomic gases \cite{Gross10}.
This figure clearly demonstrates that, using the jump operator \eqref{eq_jump},
we can obtain almost pure phase- and number- 
squeezed states.
Note that $\xi_{\rm P}$ and $\xi_{\rm N}$ of the steady states
are predicted very accurately by Eq.~\eqref{Syvariance}.

We now discuss the exact form of the steady state and show that 
our jump operator \eqref{eq_jump} drives the system into a squeezed state.  
The jump operator can be written as 
$\hat{c}=4\sqrt{\epsilon} \,e^{\chi \hat{S}_x}\hat{S}_z e^{-\chi \hat{S}_x}$, 
where $\chi=\textrm{arctanh}[(1-\epsilon)/(1+\epsilon)]$.  
If $N$ is even, one of the eigenvalues is equal to zero,  
the corresponding normalized eigenstate is 
$\phi_{\rm sq}\propto e^{\chi \hat{S}_x}|N/2\rangle$.
Here, 
we use the notation $|j\rangle$ 
for a state that has $j$ particles in mode $1$. 
State $\phi_{\rm sq}$ is a stationary state of the dynamics 
and can be written as 
$\phi_{\rm sq}=\sum_{n=-N/2}^{N/2}\alpha_n |N/2-n\rangle$,
where  
\begin{align}
\nonumber
\alpha_n=& A {N\choose N/2+ n}^{-1/2}\\
&\times\sum_{s=|n|}^{N/2}{N/2 \choose s}{N/2\choose s-|n|}
\left(\frac{1-\sqrt{\epsilon}}{1+\sqrt{\epsilon}}\right)^{2s-|n|},
\label{eq_alpha}
\end{align} 
and
$A$ is a normalization factor.
State $\phi_{\rm sq}$ is a phase-squeezed 
$(\epsilon <0)$, a number-squeezed ($\epsilon >0$), or a coherent $(\epsilon=0)$ state.
When $N$ is large, we can approximate
\begin{equation}
\alpha_{n}\simeq (2\pi \langle\Delta\hat{S}_z^2\rangle)^{-1/4} 
e^{-n^2/(4\langle\Delta\hat{S}_z^2\rangle)}.
\end{equation}
In deriving these results, $N$ was assumed to be even. 
If $N$ is odd, the steady state can be a mixed state.
However, the non-zero elements of the vector $\hat{c}\phi_{\rm sq}$
scale as $\epsilon^{(N+1)/2}$, and $\phi_{\rm sq}$ becomes
an approximate dark state of the jump operator for large $N$.
Thus, in the large $N$ limit, $\phi_{\rm sq}$ is a steady state, regardless of 
whether $N$ is even or odd.

\begin{figure}
\resizebox{8.2cm}{!}
{\includegraphics[scale=1.1]{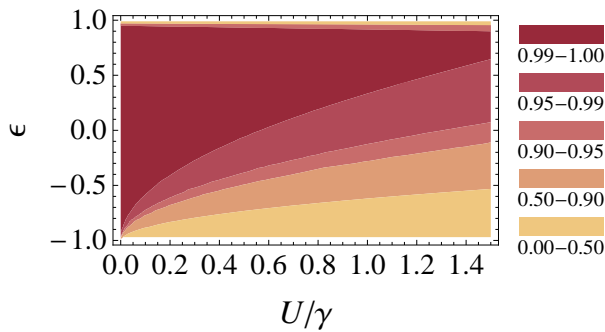}}
\caption{\label{purityregions}(Color online) Purity of the steady state 
as a function of $\epsilon$ and $U/\gamma$ for $N=100$. 
}
\end{figure}

Now, we introduce the two-site Bose-Hubbard Hamiltonian 
in addition to the dissipative terms.
For simplicity, we assume $J=0$.
We have checked numerically that the effect of non-zero $J$ 
is much less important than that of non-zero $U$ and does not 
change the results qualitatively.

The interaction term has the Fock states $|j\rangle$ as the steady states, 
while the dissipative part drives the system 
toward state $\phi_{\rm sq}$, 
which is close to the coherent state $\phi_{\rm sq}|_{\epsilon =0}$.
The steady state resulting from the competition 
between these terms is a mixed state.
The $N$ dependence of the purity and squeezing 
of the steady state decreases
with increasing $N$ and, at large $N$, these quantities are determined only  
by $\epsilon$ and $U/\gamma$. 
The disappearance of the $N$ dependence can be seen by
noting that both the interaction energy and the dissipative part
scale as $\sim N^2$.
The dashed lines in Fig.\ \ref{figure1} show the time evolution of  
the squeezing and purity for the non-zero interaction. 
We have chosen $N=100$. 
At this particle number, the $N$ dependence of the purity  of the steady 
state is already small for the values of $\epsilon$ used in this figure.
We see that it is possible to obtain 
phase- and number-squeezed states with high purity 
even for non-zero $U$, provided $U/\gamma$ is small enough.
To quantify the effectiveness of the squeezing jump operator
\eqref{eq_jump}, we plot the purity of the steady state 
at $N=100$ as a function of $\epsilon$ and $U/\gamma$ 
in Fig.\ \ref{purityregions}.
For fixed $U/\gamma$ and $|\epsilon|$, a positive $\epsilon$ leads to
larger purity than a negative one and allows us to obtain a 
high amount of number squeezing without losing purity even when $U\sim\gamma$.
This is because, for
positive $\epsilon$, the dissipative term favors number-squeezed
states that are closer to Fock states than phase-squeezed
states. Fock states are eigenstates of the interaction term and,
therefore, steady states in the absence of dissipation.

\section{Single-mode Squeezing}

Next, we consider single-mode squeezing using dissipation.
In the absence of a Hamiltonian, amplitude squeezing can be achieved using  
the jump operator,
\begin{equation}
  \hat{c} = \hat{a} + \epsilon \hat{a}^\dagger + i\lambda
\equiv \tilde{c}+i\lambda. 
\end{equation}
This is equivalent to having the jump operator $\tilde{c}$
and the Hamiltonian 
$\hat{H}=\Omega (\tilde{c}+\tilde{c}^\dagger)/2$ with 
$\Omega\equiv \gamma\lambda$.
As in the two-mode case, $\epsilon\in (-1,1)$ controls the squeezing. 
At $\epsilon=0$, 
we obtain a coherent state with the amplitude 
$\langle\hat{a}\rangle=-i\lambda$ \cite{Agarwal88}.
The equation of motion for $\langle\hat{a}\rangle$ is
\begin{equation}
\frac{\partial \langle\hat{a}\rangle}{\partial t} = -\frac{\gamma}{2} 
\left[(1-\epsilon^2)\langle\hat{a}\rangle+i\lambda(1+\epsilon)\right].
\end{equation}
We see that the time constant of $\langle\hat{a}\rangle$ is
$\tau^{-1}=\gamma(1-\epsilon^2)/2$, and  
\begin{equation}
\langle\hat{a}\rangle=-i\frac{\lambda}{1-\epsilon}\ ,
\end{equation}
in the steady state.
The equations of motion for normal-ordered correlation functions
$\langle\hat{a}^{\dagger m}\hat{a}^n\rangle$ do not contain
higher-order correlation functions and form a closed
set of equations.  Thus, the average particle numbers 
$\langle\hat{a}^\dagger\hat{a}\rangle$,
the number fluctuation $\langle\Delta\hat{n}^2\rangle
\equiv\langle(\hat{a}^\dagger\hat{a})^2\rangle
-\langle\hat{a}^\dagger\hat{a}\rangle$,
and the second-order coherence function 
$g^{(2)}\equiv\langle\hat{a}^{\dagger 2}\hat{a}^2\rangle/
\langle\hat{a}^\dagger\hat{a}\rangle^2$
for the steady state can be calculated exactly,
\begin{align}
  \langle\hat{a}^\dagger\hat{a}\rangle =& \frac{\lambda^2}{(1-\epsilon)^{2}}
+ \frac{\epsilon^2}{1-\epsilon}\ ,\\
  \langle\Delta\hat{n}^2\rangle =& \lambda^2 \frac{1+\epsilon}{(1-\epsilon)^{3}}
+ \frac{2\epsilon^2}{(1-\epsilon^2)^{2}}\ , 
\end{align}
and
\begin{equation}
  g^{(2)} = 1 + \frac{\epsilon}{\langle\hat{a}^\dagger\hat{a}\rangle^{2}}
\left[\frac{2\lambda^2}{(1-\epsilon)^{3}} 
+ \frac{\epsilon (1+\epsilon^2)}{(1-\epsilon^2)^2} \right].
\end{equation}
For $0>\epsilon\, (>\epsilon^*)$, where $\epsilon^*<0$ 
is a value such that $\langle\Delta\hat{n}^2\rangle = \lambda^2$, 
we obtain a number-squeezed state 
$\langle\Delta\hat{n}^2\rangle < \lambda^2$, which 
has a non-classical nature, characterized by $g^{(2)}<1$. 
If $\epsilon>0$, we obtain a number-anti-squeezed state 
$\langle\Delta\hat{n}^2\rangle > \lambda^2$.

\section{Lattice System}

Finally, we consider the application of our squeezing jump operator 
\eqref{eq_jump} to cold Bose gases in an optical lattice described 
by the Bose-Hubbard Hamiltonian.
A natural extension of $\hat{c}$  
for the lattice system is the following jump operator 
acting on sites $i$ and $j$:
\begin{equation}
  \hat{c}_{ij} 
= ( \hat{a}_i^\dagger + \hat{a}_{j}^\dagger )
( \hat{a}_i - \hat{a}_{j} )
+ \epsilon ( \hat{a}_i^\dagger - \hat{a}_{j}^\dagger )
( \hat{a}_i + \hat{a}_{j} ).
\label{eq_jumplat}
\end{equation}
Here, $\hat{a}_i$ is the annihilation operator at site $i$. The time evolution is given by 
 the master equation,
\begin{equation}
\frac{\partial \hat{\rho}}{\partial t}=-i[\hat{H},\hat{\rho}]+
(\gamma/2) \sum_{\langle i,j\rangle}
( 2\hat{c}_{ij}\hat{\rho}\hat{c}_{ij}^\dagger 
- \hat{c}_{ij}^\dagger\hat{c}_{ij}\hat{\rho}
- \hat{\rho}\hat{c}_{ij}^\dagger\hat{c}_{ij} ).
\end{equation}
To study the qualitative effect of the squeezing jump operator 
\eqref{eq_jumplat}, we employ the 
generalized mean-field Gutzwiller approach \cite{Diehl10,Tomadin10}. 
It consists of a product ansatz for the density operator
$\hat{\rho}=\bigotimes_i \hat{\rho}_i$, where 
$\hat{\rho}_i\equiv \textrm{Tr}_{\ne i}[\hat{\rho}]$ 
are the reduced local density operators for site $i$
and the site-decoupling approximation 
$\hat{H}=\sum_i\hat{h}_i$, with 
\begin{equation}
\hat{h}_i=-J\sum_{\langle i'|i\rangle} 
(\langle\hat{a}_{i'}\rangle\hat{a}_i^\dagger 
+ \langle\hat{a}_{i'}^\dagger\rangle\hat{a}_i) -\mu \hat{n}_i
+ (U/2)\hat{n}_i(\hat{n}_i-1)
\end{equation}
and $\hat{n}_i\equiv\hat{a}^\dagger_i\hat{a}_i$.

We are interested in the region of the higher filling factor 
$\bar{n}\equiv\langle\hat{n}\rangle \gtrsim 3$ 
where the filling factor dependence becomes small and 
we can obtain a universal result.  
Besides, it has been shown that 
distinct commensurability effects are absent even 
at low $\bar{n}$ \cite{Diehl10,Tomadin10}.
In the following calculations, we consider a homogeneous system 
(hereafter, we omit the site index $i$) with 
$\bar{n}=4$ as an example.
Because of the large $\bar{n}$, we expect that 
the generalized mean-field Gutzwiller approach is quantitatively reliable
for higher dimensions.
We choose a pure coherent state as the initial state of 
the local density operator
and study the resulting steady state of the time evolution.

\begin{figure}[tb]
\resizebox{8.2cm}{!}
{\includegraphics{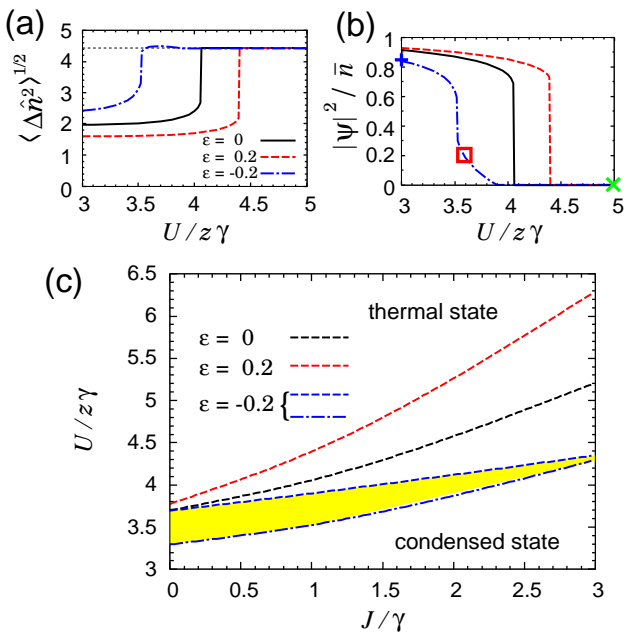}}
\caption{\label{fig_lat}(Color online) 
Steady states of the dissipative dynamics in an optical lattice 
driven by the squeezing jump operator \eqref{eq_jumplat}.
We have set $\bar{n}=4$. Panels (a) and (b) show the number fluctuation 
$\langle \Delta\hat{n}^2\rangle^{1/2}$
and the condensate fraction $|\psi|^2/\bar{n}$, respectively, for $J/\gamma=1$. 
In (b), the blue plus sign, red box, and green cross correspond to 
the parameter values used in Figs.\ \ref{fig_rho}(b), (c),
and (d), respectively.  
Panel (c) gives the non-equilibrium phase diagram.
The dashed lines show the boundary
between the phases with $|\psi|^2=0$ (thermal state) and $|\psi|^2\ne0$. 
The phase with a thermal-like condensed state obtained
for $\epsilon=-0.2$ is shown by the filled yellow region.
}
\end{figure}

Figure \ref{fig_lat}(a) shows the local number fluctuation
$\langle\Delta\hat{n}^2\rangle^{1/2}\equiv 
(\langle\hat{n}^2\rangle-\bar{n}^2)^{1/2}$ 
as a function of $U/z\gamma$ ($z$ is the coordination number)
for $J/\gamma=1$ and $\epsilon=-0.2$, 0, and 0.2.
As in the two-mode case, positive (negative) $\epsilon$ yields smaller 
(larger) $\langle\Delta\hat{n}^2\rangle^{1/2}$ corresponding to
number squeezing (anti-squeezing). 
In Fig.\ \ref{fig_lat}(b), we show the condensate fraction 
$|\psi|^2/\bar{n}$ of the steady state. Here, 
 $\psi\equiv \langle\hat{a}\rangle$ is the order parameter.
Similar to the two-mode system, the interaction term 
favors states with a definite number of particles and, thus,
suppresses the off-diagonal order. Hence, 
$|\psi|^2/\bar{n}$ decreases monotonically  with increasing $U$
and finally vanishes.
The boundary between the phase with zero and non-zero $|\psi|^2/\bar{n}$
is shown by the dashed lines in Fig.\ \ref{fig_lat}(c) 
for various values of $J$.
We find that the region of the condensed phase $|\psi|^2/\bar{n}> 0$ 
grows (shrinks) for positive (negative) $\epsilon$ \cite{note_reason}. 
This allows us to control the phase boundary of the non-equilibrium
phase diagram Fig.\ \ref{fig_lat}(c) by changing the value of $\epsilon$.

\begin{figure}[t]
\resizebox{8.2cm}{!}
{\includegraphics{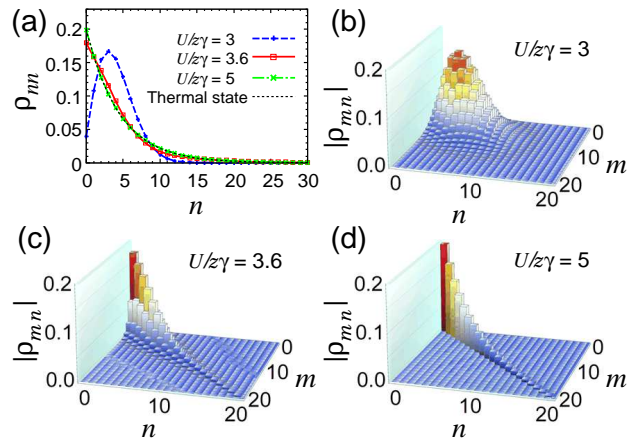}}
\caption{\label{fig_rho}(Color online) 
Matrix element $\rho_{mn}$ of the local density operator $\hat{\rho}$ 
of the steady state driven by the squeezing jump operator
\eqref{eq_jumplat} with negative $\epsilon$.
Here, we set $\epsilon=-0.2$, $\bar{n}=4$, and $J/\gamma=1$.
Panel (a) shows the particle-number statistics.
Panels (b), (c), and (d) show $|\rho_{mn}|$ of the condensed state,
thermal-condensed state, and thermal state, respectively.
}
\end{figure}

We note that 
the behavior of $|\psi|^2/\bar{n}$ 
depends strongly on whether $\epsilon <0$ or $\epsilon \geq 0$
[see Fig.\ \ref{fig_lat}(b)]. 
In the latter case, $|\psi|^2/\bar{n}$ decreases suddenly  to zero 
(solid and dashed lines)
while in the case of negative $\epsilon$, 
$|\psi|^2/\bar{n}$ first drops to a non-zero value 
and then gradually decreases to zero with further increasing $U/z\gamma$
(dashed-dotted line).
To better understand the properties of the
state with a small but non-zero $|\psi|^2/\bar{n}$,
we show the matrix elements $\rho_{mn}$ (in the Fock-state basis) 
of the steady-state 
density operator $\hat{\rho}$ at $U/z\gamma=3.6$
for $\epsilon=-0.2$ and $J/\gamma=1$ in Fig.\ \ref{fig_rho}(c).
For comparison, we also show $\rho_{mn}$ for the
condensed state with large $|\psi|^2/\bar{n}$ at $U/z\gamma=3$ and
a state with $|\psi|^2/\bar{n}=0$ at $U/z\gamma=5$.  
The state with $|\psi|^2/\bar{n}=0$ is well described
by the thermal state $\rho_{nn}=\bar{n}^n/(\bar{n}+1)^{n+1}$, 
which does not have off-diagonal elements 
[see Figs.\ \ref{fig_rho}(a) and \ref{fig_rho}(d)].
Figure \ref{fig_rho}(a) clearly shows that the diagonal elements $\rho_{nn}$  
at $U/z\gamma=3.6$ are very close to $\rho_{nn}$ of the thermal state.
Simultaneously, the steady state at $U/z\gamma=3.6$ has non-zero
off-diagonal elements [Fig.\ \ref{fig_rho}(c)].
We call such a state the thermal-condensed state. It is characterized by  
almost thermal particle-number statistics and a non-zero condensate fraction.
We also note that the thermal-condensed state has slightly
larger $\langle\Delta\hat{n}^2\rangle^{1/2}$ than the thermal state 
[see Fig.\ \ref{fig_lat}(a)].
The thermal-condensed state appears as a new phase in the
non-equilibrium steady-state phase diagram for $\epsilon <0$. 
It is represented by the 
yellow region in Fig.\ \ref{fig_lat}(c). 
This phase is separated from the
ordinary condensed phase by a jump in the order parameter
[see dashed-dotted line in Fig.\ \ref{fig_lat}(b)]
and it differs from the thermal phase by
having a non-zero order parameter.

\section{Conclusion}
We have proposed a way to produce number- and
phase-squeezed states in a two-mode Bose system using dissipation.
When applied to an optical lattice, our scheme can be used to control
the phase boundaries of the steady-state phase diagram. It also allows
us to realize a new phase characterized by a non-zero condensate
fraction and thermal-like particle-number statistics.

\begin{acknowledgments}
We are grateful to S. Diehl, C. J. Pethick, K.-A. Suominen, and T. Takimoto 
for helpful discussions and comments.
G.W. acknowledges the Max Planck Society, 
the Korea Ministry of Education, Science and Technology,
Gyeongsangbuk-Do, and Pohang City for the support of the JRG at APCTP.
\end{acknowledgments}

\appendix

\section{Expectation values of $\hat{S}_x$ and $\hat{S}_x^2$ and Bogoliubov backreaction formalism \label{sec_append}}

\begin{figure}[t!]
\resizebox{4.cm}{!}
{\includegraphics{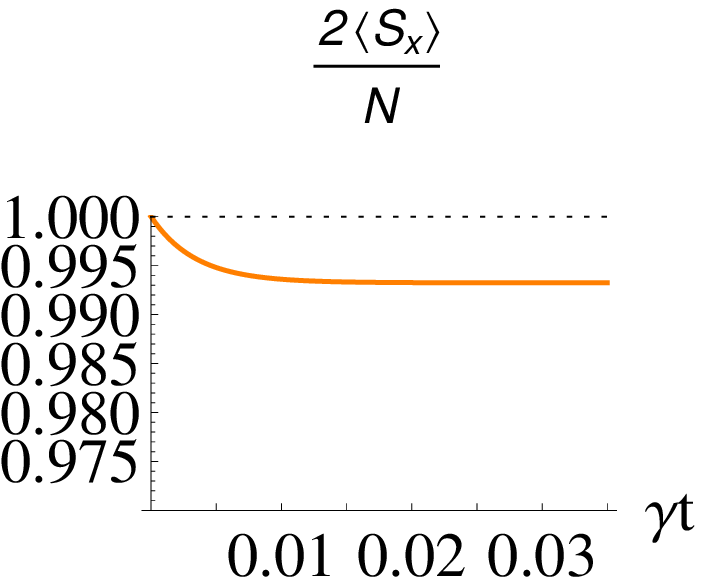}}
\resizebox{4.cm}{!}
{\includegraphics{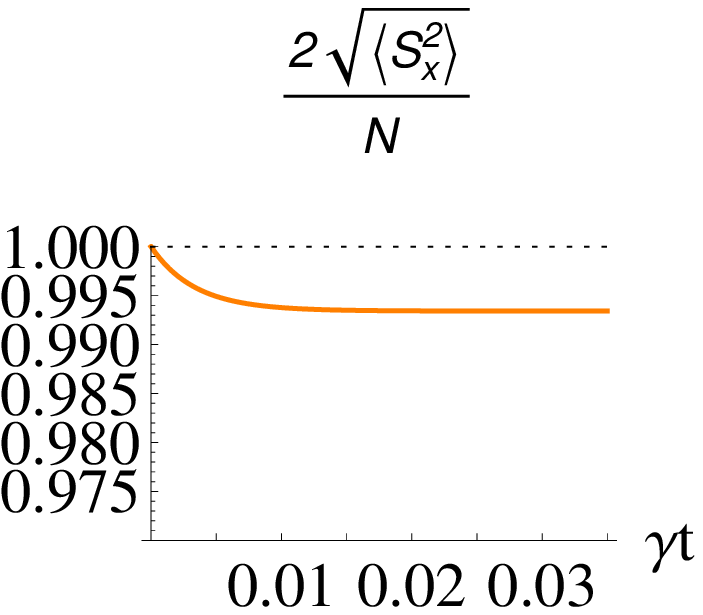}}
\vspace{1cm}\\
\resizebox{4.cm}{!}
{\includegraphics{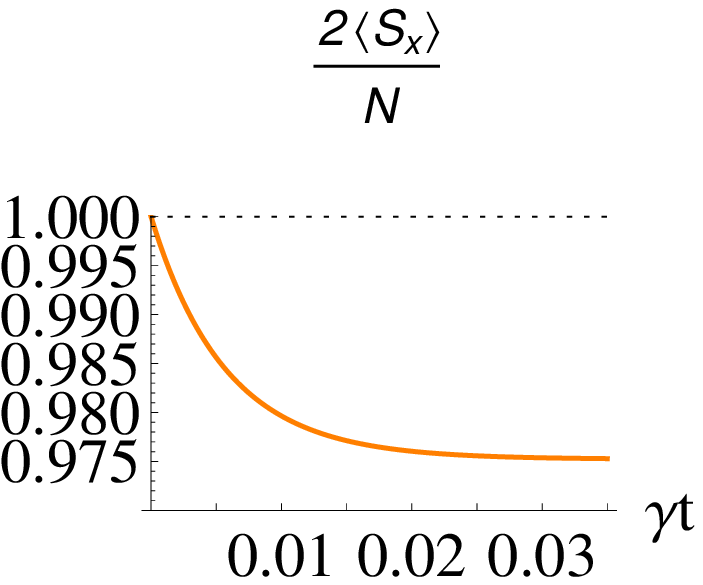}}
\resizebox{4.cm}{!}
{\includegraphics{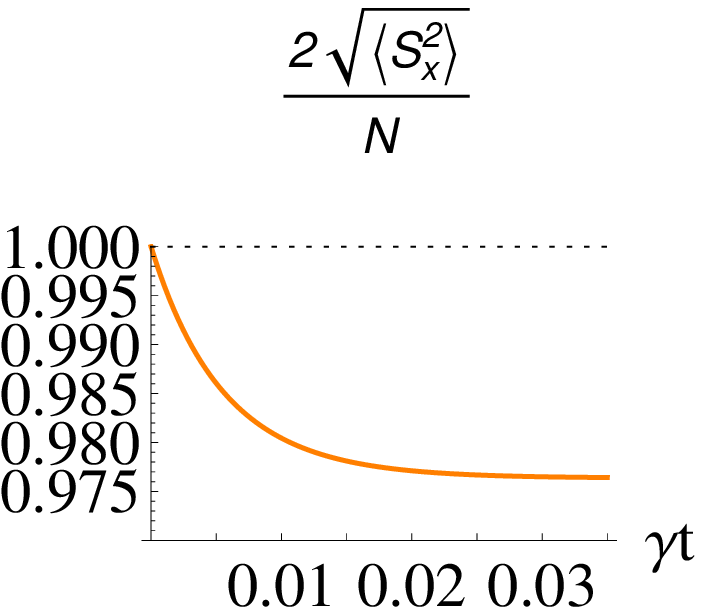}}
\caption{(Color online) The ratio of the exact values of $\langle\hat{S}_x\rangle$
  and $\sqrt{\langle\hat{S}_x^2\rangle}$ to their approximate value,
  $N/2$.  The parameters are the same as for the solid lines in Fig.\ \ref{figure1}: $N=100$, $J=0$, and $U/\gamma=0$.  The upper 
  panels correspond to $\epsilon=-0.50$, and the bottom ones correspond to
  $\epsilon=0.74$.
\label{fig1}}
\end{figure}

\begin{figure}[t!]
\resizebox{4.cm}{!}
{\includegraphics{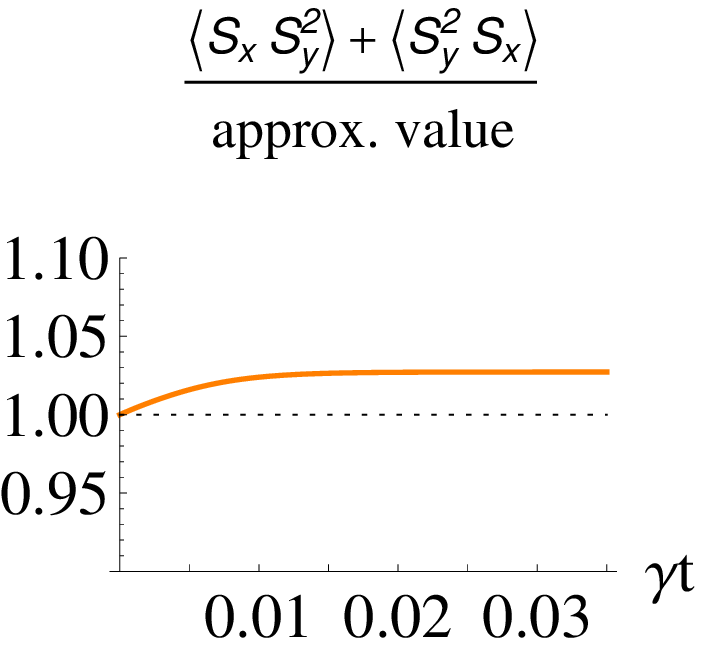}}
\resizebox{4.cm}{!}
{\includegraphics{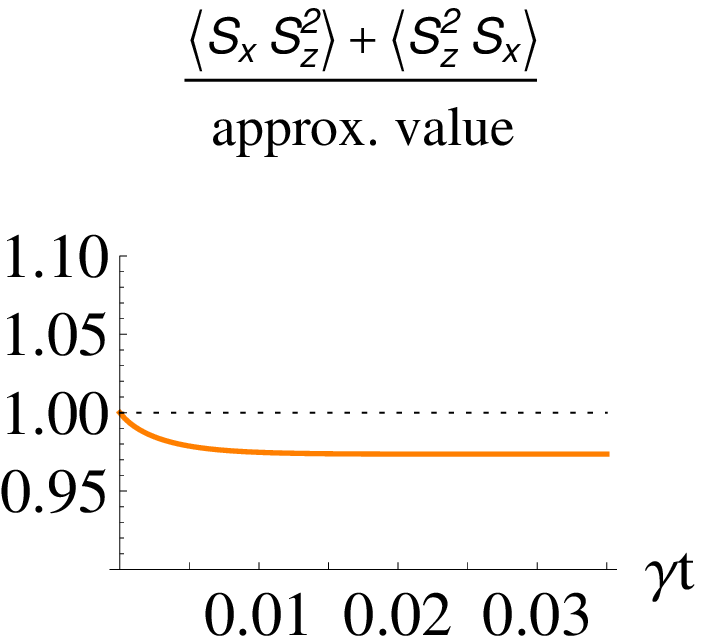}}
\vspace{1cm}\\
\resizebox{4.cm}{!}
{\includegraphics{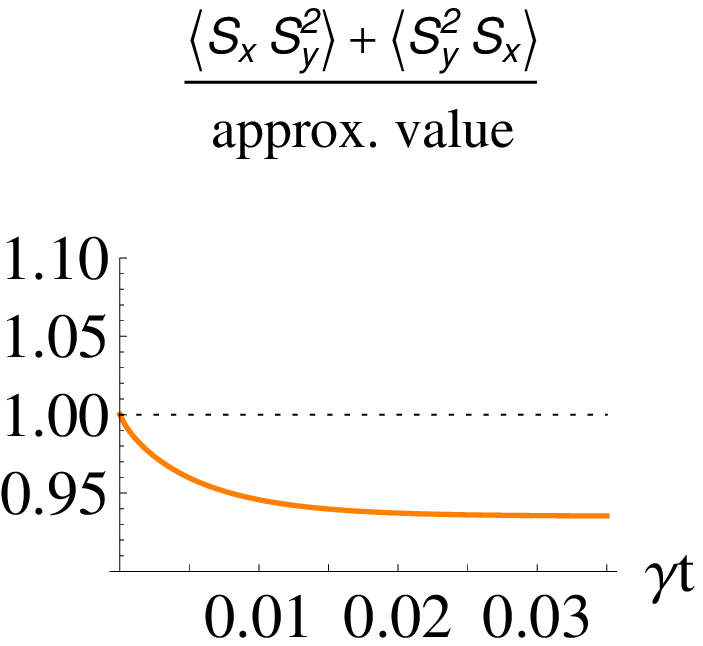}}
\resizebox{4.cm}{!}
{\includegraphics{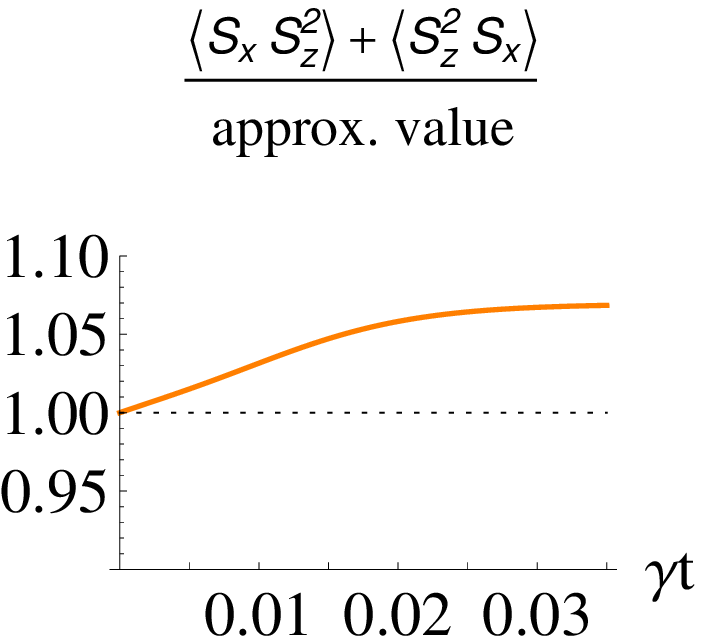}}
\caption{(Color online) The ratio of the exact values of
  $\langle\hat{S}_x\hat{S}_y^2\rangle+\langle\hat{S}_y^2\hat{S}_x\rangle$
  and
  $\langle\hat{S}_x\hat{S}_z^2\rangle+\langle\hat{S}_z^2\hat{S}_x\rangle$
  to their approximate values given by Eqs.\ (\ref{sxsysy}) and
  (\ref{sxszsz}), respectively.  The parameters are the same as for the
  solid lines in Fig.\ \ref{figure1}: $N=100$, $J=0$, and
  $U/\gamma=0$.  The upper panels correspond to $\epsilon=-0.50$, and
  the lower ones correspond to $\epsilon=0.74$.
\label{fig2}}
\end{figure}

In the derivation of Eq.\ (\ref{Syvariance}), we have used the following 
approximation:
\begin{equation}
\langle\hat{S}_x\rangle, \sqrt{\langle\hat{S}_x^2\rangle} \simeq \frac{N}{2}. 
\label{coherent}
\end{equation} 
In order to illustrate the validity of this approximation, we plot the
ratio of the exact and approximate expectation values in
Fig.\ \ref{fig1}.  The values of the parameters are the same as in
Fig.\ \ref{figure1}.  Note that the approximation
(\ref{coherent}) holds very well, and the relative error is less than
$1\%$ for $\epsilon=-0.5$ (upper panels) and 
less than $2.5\%$ for $\epsilon=0.74$ (lower panels)
throughout the time evolution.  We have observed that the
error of the approximation (\ref{coherent}) for the final steady
states scales roughly as $\sim 1/N$ and, thus, becomes even smaller for
larger numbers of particles.

Another approximation used in our article was based on the so-called
Bogoliubov backreaction formalism \cite{Vardi01}, which approximates the expectation
value of a product of three operators as
\begin{align}
\langle \hat{S}_i \hat{S}_j \hat{S}_k\rangle
\simeq& \langle \hat{S}_i\hat{S}_j\rangle \langle \hat{S}_k\rangle
+ \langle \hat{S}_i\rangle \langle \hat{S}_j\hat{S}_k\rangle 
+ \langle \hat{S}_i\hat{S}_k\rangle \langle \hat{S}_j\rangle\nonumber\\
& -2 \langle \hat{S}_i\rangle \langle \hat{S}_j\rangle \langle \hat{S}_k\rangle.
\label{bbr}
\end{align}
In the derivation of $\langle\Delta \hat{S}_y^2\rangle$ in Eq.\ (\ref{Syvariance}), we used
\begin{align}
\langle \hat{S}_x \hat{S}_y^2\rangle + \langle \hat{S}_y^2\hat{S}_x\rangle
\simeq& 2(\langle \hat{S}_x\hat{S}_y\rangle +\langle \hat{S}_y\hat{S}_x\rangle)\langle \hat{S}_y\rangle
+ 2\langle \hat{S}_x\rangle \langle \hat{S}_y^2\rangle \nonumber\\
& -4 \langle \hat{S}_x\rangle \langle \hat{S}_y\rangle^2,
\label{sxsysy}
\end{align}
and, in the derivation of $\langle\Delta \hat{S}_z^2\rangle$, we used
\begin{align}
\langle \hat{S}_x \hat{S}_z^2\rangle + \langle \hat{S}_z^2\hat{S}_x\rangle
\simeq& 2(\langle \hat{S}_x\hat{S}_z\rangle +\langle \hat{S}_z\hat{S}_x\rangle)\langle \hat{S}_z\rangle
+ 2\langle \hat{S}_x\rangle \langle \hat{S}_z^2\rangle \nonumber\\
& -4 \langle \hat{S}_x\rangle \langle \hat{S}_z\rangle^2.
\label{sxszsz}
\end{align}

In Fig. \ref{fig2}, we plot the exact values of 
$\langle \hat{S}_x \hat{S}_y^2\rangle + \langle \hat{S}_y^2\hat{S}_x\rangle$ and 
$\langle \hat{S}_x \hat{S}_z^2\rangle + \langle \hat{S}_z^2\hat{S}_x\rangle$
divided by their approximate values given by Eqs.\ (\ref{sxsysy}) and (\ref{sxszsz}),
respectively.
The values of the parameters are the same as those in Fig.\ \ref{figure1}.  
We see that the relative error is less than $3\%$ for $\epsilon=-0.50$
(upper panels) and is less than $7\%$ for $\epsilon=0.74$ (lower panels).
As for $\langle\hat{S}_x\rangle$ and
$\sqrt{\langle\hat{S}_x^2\rangle}$, the error of the approximations (\ref{sxsysy}) and (\ref{sxszsz}) for the final
steady states scales roughly  as $\sim 1/N$, which justifies the use 
of the approximation for large $N$ as in the case of Fig.\ \ref{figure1}.

Finally, we note that the approximations are valid for the final
states of the time evolution regardless of the initial state.  In the
calculations of Fig.\ \ref{figure1}, we chose the coherent
state as the initial state.  The system is, however, driven to the
squeezed states no matter how the initial state is chosen [the proof
has been given in the paragraph of Eq.\ (\ref{eq_alpha}) 
in Sec. \ref{2mode}]. The squeezed states satisfy the approximate
equations (\ref{coherent}) and (\ref{sxsysy})-(\ref{sxszsz}) to the
accuracy shown in Figs.\ \ref{fig1} and \ref{fig2}.


\begin{thebibliography}{99}
%
\bibitem{Diehl08} S. Diehl, A. Micheli, A. Kantian, B. Kraus, H. P. B\"uchler, and P. Zoller,
  Nature Phys. {\bf 4}, 878 (2008).
%
\bibitem{Kraus08} B. Kraus, H. P. B\"uchler, S. Diehl, A. Kantian, A. Micheli, and P. Zoller,
  Phys.\ Rev.\ A {\bf 78}, 042307 (2008).
%
\bibitem{Verstraete09} F. Verstraete, M. M. Wolf, and J. I. Cirac,
  Nature Phys. {\bf 5}, 633 (2009).
%
\bibitem{Witthaut08} D. Witthaut, F. Trimborn, and S. Wimberger, 
  Phys.\ Rev.\ Lett. {\bf 101}, 200402 (2008).
%
\bibitem{GarciaRipoll09} J. J. Garc\'ia-Ripoll, S. D\"urr, N. Syassen, D. M. Bauer, M. Lettner, G. Rempe, and J. I. Cirac,
  New J.\ Phys. {\bf 11}, 013053 (2009).
%
\bibitem{Diehl10} S. Diehl, A. Tomadin, A. Micheli, R. Fazio, and P. Zoller,
  Phys.\ Rev.\ Lett. {\bf 105}, 015702 (2010).
%
\bibitem{Weimer10} H. Weimer, M. M\"uller, I. Lesanovsky, P. Zoller, and H. P. B\"uchler,
  Nature Phys. {\bf 6}, 382 (2010). 
%
\bibitem{Diehl10b} S. Diehl, W. Yi, A. J. Daley, and P. Zoller,
  Phys.\ Rev.\ Lett. {\bf 105}, 227001 (2010).
%
\bibitem{Tomadin10} A. Tomadin, S. Diehl, and P. Zoller,
  Phys.\ Rev.\ A {\bf 83}, 013611 (2011).
%
\bibitem{Krauter10} H. Krauter, C. A. Muschik, K. Jensen, W. Wasilewski, J. M. Petersen, J. I. Cirac, and E. S. Polzik
  Phys.\ Rev.\ Lett. {\bf 107}, 080503 (2011).
%
\bibitem{Barreiro10} J. T. Barreiro, P. Schindler, O. G\"uhne, T. Monz, M. Chwalla, C. F. Roos, M. Hennrich, and R. Blatt,
  Nature Phys. {\bf 6}, 943 (2010); 
  {\it ibid.} Nature {\bf 470} 486 (2011).
%
\bibitem{dynamical} A. Eckardt, C. Weiss, and M. Holthaus,
  Phys.\ Rev.\ Lett. {\bf 95}, 260404 (2005);
  C. E. Creffield, Phys.\ Rev.\ Lett. {\bf 99}, 110501 (2007);
  F. Piazza, L. Pezz\'e, and A. Smerzi, 
  Phys.\ Rev.\ A {\bf 78}, 051601(R) (2008);
  G. Watanabe,
  Phys.\ Rev.\ A {\bf 81}, 021604(R) (2010);
  C. Ottaviani, V. Ahufinger, R. Corbal\'an, and J. Mompart,
  Phys.\ Rev.\ A {\bf 81}, 043621 (2010);
  M. S. Rudner, L. M. K. Vandersypen, V. Vuleti\'c, and L. S. Levitov,
  Phys.\ Rev.\ Lett. {\bf 107}, 206806 (2011);
  M. A. Leung, K. W. Mahmud, and W. P. Reinhardt, e-print arXiv:1006.2556.
%
\bibitem{Esteve08} J. Est\`eve, C. Gross, A. Weller, S. Giovanazzi, and M. K. Oberthaler,
  Nature {\bf 455}, 1216 (2008).
%
\bibitem{Appel09} J. Appel, P. J. Windpassinger, D. Oblak, U. B. Hoff, N. Kj{\ae}rgaard, and E. S. Polzik,
  Proc.\ Natl.\ Acad.\ Sci.\ USA {\bf 106}, 10960 (2009). 
%
\bibitem{Cronin09} A. Cronin, J. Schmiedmayer, and D. E. Pritchard,
  Rev.\ Mod.\ Phys. {\bf 81}, 1051 (2009).
%
\bibitem{Leroux10} I. D. Leroux, M. H. Schleier-Smith, and V. Vuleti\'c, 
  Phys.\ Rev.\ Lett. {\bf 104}, 073602 (2010).
%
\bibitem{Gross10} C. Gross, T. Zibold, E. Nicklas, J. Est\`eve, and M. K. Oberthaler,
  Nature {\bf 464}, 1165 (2010).
%
\bibitem{Grond10} J. Grond, U. Hohenester, I. Mazets, and J. Schmiedmayer,
  New J.\ Phys. {\bf 12}, 065036 (2010).
%
\bibitem{Lee11} C. Lee, J. Huang, H. Deng, H. Dai, and J. Xu,
  Front.\ Phys. {\bf 7}, 109 (2012).
%
\bibitem{note_nonzerophi} It is straightforward to generalize the squeezing 
jump operator \eqref{eq_jump} to allow for a non-zero relative phase $\phi$\,,
$\hat{c}(\phi)=2(1+\epsilon)\hat{S}_z
-2i(1-\epsilon)(\cos{\phi}\hat{S}_y+\sin{\phi}\hat{S}_x)$.
%
\bibitem{squeeze2} G. Watanabe, H. M\"akel\"a, and S. Diehl (to be published).

\bibitem{note_implement} There we consider two narrow wells
embedded in a wide harmonic potential.
Each narrow well holds either state $\phi_1$ or state $\phi_2$, 
corresponding to $\hat{a}_1$ and $\hat{a}_2$.
States $\phi_1$ and $\phi_2$ are Raman coupled to an even-parity
state $\phi_e$ 
(with the Rabi frequencies $\Omega_1$ and $-\Omega_1$, respectively)   
and an odd-parity one $\phi_o$ (with equal Rabi frequency $\Omega_2$)
in the wide harmonic potential.  Atoms excited to $\phi_e$ and $\phi_o$
decay into $\phi_1$ and $\phi_2$ by emitting Bogoliubov excitations
in the background Bose-Einstein condensate.
%
\bibitem{Kitagawa93} M. Kitagawa and M. Ueda, 
  Phys.\ Rev.\ A {\bf 47}, 5138 (1993).
%
\bibitem{note_approx} 
See the Appendix \ref{sec_append} for a more detailed discussion on 
the validity of the approximation.
%
\bibitem{Vardi01} A. Vardi and J. R. Anglin,
  Phys.\ Rev.\ Lett. {\bf 86}, 568 (2001).
%
\bibitem{note_bbr} 
For the steady states of our setup, the relative
error of this approximation used in the derivation of Eq.\ (\ref{Syvariance}) 
scales roughly as $1/N$.
%
\bibitem{Agarwal88} G. S. Agarwal,
  J.\ Opt.\ Soc.\ Am.\ B {\bf 5}, 1940 (1988).
%
\bibitem{note_reason} 
At first sight, this is somewhat counterintuitive because
$\epsilon<0$ suppresses local phase fluctuations.  This result can be
understood by noting that, in the steady state, $|\rho_{n+i,n-i}|<\rho_{nn}$
for any integer $i$ such that $-n\leq i\leq n$ (see Fig. \ref{fig_rho}).  
In the number-squeezed state, the distribution of the diagonal elements
$\rho_{nn}$ becomes more peaked. This makes having larger
off-diagonal elements possible and enhances the condensate fraction.  Number
anti-squeezing results in the opposite behavior.
%
\end{thebibliography}
\end{document}